# Spatial quantum noise interferometry in expanding ultracold atom clouds

Simon Fölling, Fabrice Gerbier, Artur Widera, Olaf Mandel, Tatjana Gericke & Immanuel Bloch

*Institut für Physik, Johannes Gutenberg-Universität, Staudingerweg 7, D-55099 Mainz, Germany*

**In a pioneering experiment[1], Hanbury Brown and Twiss (HBT) demonstrated that noise correlations could be used to probe the properties of a (bosonic) particle source through quantum statistics; the effect relies on quantum interference between possible detection paths for two indistinguishable particles. HBT correlations—together with their fermionic counterparts[2–4]—find numerous applications, ranging from quantum optics[5] to nuclear and elementary particle physics[6]. Spatial HBT interferometry has been suggested[7] as a means to probe hidden order in strongly correlated phases of ultracold atoms. Here we report such a measurement on the Mott insulator[8–10] phase of a rubidium Bose gas as it is released from an optical lattice trap. We show that strong periodic quantum correlations exist between density fluctuations in the expanding atom cloud. These spatial correlations reflect the underlying ordering in the lattice, and find a natural interpretation in terms of a multiple-wave HBT interference effect. The method should provide a useful tool for identifying complex quantum phases of ultracold bosonic and fermionic atoms[11–15].**

Although quantum noise correlation analysis is now a basic tool in various areas of physics, applications to the field of cold atoms have been scarce. Most of these concentrate on photon correlation techniques from quantum optics[5,16]. It was not until 1996 that bunching of cold (but non-degenerate) bosonic atom clouds could be directly measured[17], followed by the observation of reduced inelastic losses due to a modification of local few-body correlations by quantum degeneracy[18–20].

In our experiment, we directly measure the spatial correlation function of the density fluctuations in a freely expanding atomic cloud[7,21–23]. We create an ultracold Bose gas in an optical lattice with several hundred thousand occupied lattice sites, and record the density distribution after sudden switch-off of the trapping potential and a fixed period of free expansion (the 'time of flight'). Resonant absorption of a probe laser[24] yields the two-dimensional column density of the cloud, that is, the density profile integrated along the probe line of sight, as illustrated in Fig. 1a. It should be noted that the density after the time of flight reflects the in-trap momentum distribution rather than the in-trap density distribution. We performed the experiment with a Bose gas initially in the Mott insulator regime[8–10], where repulsive interactions pin the atomic density to exactly an integer number of atoms per lattice site, typically between one and three. In this Mott insulator phase, the average density distribution after expansion is simply given by the incoherent sum of all single particle wavefunctions released from each lattice site—a featureless gaussian. However, a typical single shot absorption image as shown in Fig. 2a and b exhibits large fluctuations around this average. It is the purpose of this Letter to demonstrate that these fluctuations are related to intrinsic quantum noise and that their HBT-type correlations contain information on the spatial order in the lattice that is absent from the average density.

To analyse the fluctuations, we introduce the spatially averaged, normalized density–density correlation function

$$C(\mathbf{d}) = \frac{\int \langle n(\mathbf{x}+\mathbf{d}/2) \cdot n(\mathbf{x}-\mathbf{d}/2) \rangle d^2\mathbf{x}}{\int \langle n(\mathbf{x}+\mathbf{d}/2) \rangle \langle n(\mathbf{x}-\mathbf{d}/2) \rangle d^2\mathbf{x}} \quad (1)$$

which denotes the conditional probability of finding two particles at two positions separated by a vector **d**, averaged over all such positions. In equation (1), $n(\mathbf{x})$ is the column density obtained from a single absorption image and the brackets $\langle \ \rangle$ denote averaging over an ensemble of independently acquired images. Uncorrelated particles correspond to $C(\mathbf{d})=1$, whereas $C(0)>1$ indicates a tendency of particles to bunch, typical for bosons. In Fig. 2c and d, an experimentally obtained correlation function is shown. In striking contrast to the atomic density distribution of Fig. 2a, sharp peaks emerge. They appear on positions corresponding to the reciprocal lattice vectors of the original periodic trapping potential.

We found the correlation patterns to be robust in the Mott insulating regime, and observed them over a broad range of lattice depths. For a planar lattice of several thousand one-dimensional decoupled Bose gases with random phases[10,25], we observed similar density correlations. The latter case is related to a recent experiment[26], where single shot interference patterns were observed from 30 independent Bose–Einstein condensates with random phases. Both these cases can be described through a classical field model, whereas the case of a Mott insulator presented here requires a full quantum treatment and detection of the atom number distribution at the atomic shot noise level.

In order to explain the origin of the correlations in the density fluctuations and their regularity, let us first consider the simple model illustrated in Fig. 1b. Two bosonic atoms in a periodic potential are initially localized at two lattice sites $i$ and $j$ separated by $n_{ij}$ lattice spacings. When these particles are released from the trapping potential, one can show[7,21] that the joint detection probability by two detectors separated by a distance $d$ is sinusoidally varying as a function of $d$ (Fig. 1c). The spatial wavevector $2\pi n_{ij}/l$ of this modulation is determined by the separation of the sources and the characteristic length,

$$l = \frac{h}{ma_{lat}}t \quad (2)$$

where $t$ denotes the time of flight, $h$ is Planck's constant, $m$ is the atomic mass, and $a_{\text{lat}}$ is the lattice spacing. For a large number of perfectly distributed but independent sources corresponding to the individual sites of the Mott insulator, the joint detection amplitudes for all possible pairs in the source have to be added. The wavenumbers associated with every pair formed from such a distribution are then integer multiples of $2\pi/l$, so that away from $d=0$ the amplitudes add up constructively wherever the detector separation is a multiple of $l$. A lattice pattern of sharp peaks reproducing the reciprocal lattice therefore emerges as the number of particles increases, each peak width being roughly determined by $l/N_s$, with $N_s$ being the number of occupied sites in the lattice in one dimension. Owing to the different quantum statistics of bosons and fermions the fundamental sine components have opposite signs, and therefore positive peaks emerge for bosons and negative peaks for fermions.

In the experiments, each detector in the above model is represented by a pixel of the CCD (charge-coupled device) camera, which detects the absorption image of the atom cloud. The array of pixels in this camera thereby samples the column density of the atomic cloud, according to $n=N_{bin}/A_{px}$. Here $N_{bin}$ is the number of atoms detected within a column defined by the area $A_{px}$ of each pixel and the direction of propagation of the probe light. In addition, each bin is smoothed by the point spread function of our imaging system, which we approximate by a gaussian. In our case, its root mean square (r.m.s.) radius $\sigma \approx 5.6$ µm is larger than the pixel size (4.4 µm) and determines an effective width for each bin.

The correlation for such a coarse-grained density can still be calculated from equation (1) by taking the integration over the probe line of sight into account and convoluting the resulting correlation signal with the resolution function described above (see also Methods section). The outcome of this calculation can be qualitatively understood from statistical considerations. One expects the amount of relative fluctuations in a single detection bin to scale as $1/\sqrt{N_{bin}}$ for $N_{bin} \gg 1$, implying a $1/N_{bin}$ scaling for the amplitude increase above the uncorrelated level $C(\mathbf{d})=1$ (see equation (1)). However, in our case this signal is distributed over $N_{peaks} \approx 4\pi(w/l)^2$ correlation peaks within the expanding density envelope of width $w$. To obtain the magnitude of the signal, we calculate the average number of atoms per bin as $N_{bin}=N_{atoms}(\sigma/w)^2$, with $N_{atoms}$ representing the total atom number. This yields a correlation amplitude per peak:

$$C(\mathbf{d}_{pk}) - 1 \approx \frac{1}{N_{peaks}} \frac{1}{N_{bin}} \approx \frac{1}{4\pi N_{atoms}} \left(\frac{l}{\sigma}\right)^2 \qquad (3)$$

The estimate in equation (3) agrees with a rigorous calculation assuming a homogeneous Mott insulator with unity filling (see Methods). For typical parameters of the experiments ($5\times10^5$ atoms, $l/\sigma=40$), it yields a correlation amplitude of $\sim 3\times10^{-4}$, in agreement with our observations.

To confirm our analysis, we plot in Fig. 3 the experimental correlation signal from the Mott insulator versus expansion time and atom number. This signal is defined by the volume under the lateral peaks, that is, the product of the peak height times its area as determined by a gaussian fit. The resulting 'correlation signal' does not require a precise determination of the resolution, and is rather insensitive to defocusing or calibration errors. According to equation (3), it depends quadratically on the time of flight and inversely on atom number for homogeneous filling. However, for a Mott insulator in a harmonic trap, a shell structure develops for increasing filling[8,27]. A model of this atom number distribution (see Methods) predicts a reduction of the $(N_{atoms})^{-1}$ scaling behaviour for large atom numbers to $(N_{atoms})^{-0.64}$. Using a combined fit to both data sets in Fig. 3a and b, the measured exponent of atom number scaling is 0.78±0.15, close to the expected value. However, the amplitude of the signal is 40% lower than what would be expected from our simple theoretical model.

For a Bose–Einstein condensate in an optical lattice, a flat spatial correlation function is expected[28]. Obtaining the correlation function in this regime, however, turned out not to be experimentally possible. In the actual experiment small fluctuations of the superfluid interference pattern between the individual images exist, owing to technical reasons such as shot-to-shot atom number variations, or excitations of the condensate by external perturbations. Such fluctuations are not cancelled by the normalization in equation (1), and the associated correlations turn out to be stronger than the quantum noise correlations observed in the Mott insulating case. We attribute the much more robust quantum noise correlation signal of a Mott insulator to the gapped excitation spectrum (with an energy gap ~3 kHz in our case), which protects this state from external perturbations that would otherwise degrade the correlations. We also investigated the case of a thermal cloud significantly above condensation temperature but with no observable population in the excited Bloch bands, for which we did not find a correlation signal. A possible explanation for this could be the decrease in signal (compared to the Mott insulating state) due to an increased spatial size of the system in combination with an increased noise background due to density and temperature fluctuations. We have considered the possibility that the correlation signal we observed in the Mott insulating case could be produced by shot-to-shot fluctuations of a residual fraction of atoms with long range coherence. In order to rule out this effect, we have checked that the regions that would contain the peaks of the diffraction pattern can be excluded from the analysis without significantly affecting the noise correlation pattern.

In conclusion, we have demonstrated that spatial quantum noise correlations in expanding atom clouds can be used to reveal the ordering of indistinguishable particles in optical lattices. They enable the direct and easy detection of many of the more complex and intriguing quantum phases that have been predicted for ultracold bosonic and fermionic atoms—for example, antiferromagnets or spin-waves in two-component spin mixtures loaded into the optical lattice[12,13]. Antiferromagnetic ordering or charge density waves in Fermi gases or Bose–Fermi mixtures[14,15], for example, would yield additional correlation peaks at momenta given by half the reciprocal lattice vector[7].

We note that after submission of this manuscript, we received a preprint[29] reporting the use of noise correlations in expanded atom clouds for identifying the fragments produced by ultracold molecule dissociation.

## Methods

**Experimental sequence**

Atomic Mott insulators are prepared by loading a Bose–Einstein condensate of up to $6\times10^5$ atoms of $^{87}$Rb into an optical lattice potential. For this, three optical standing waves of wavelength $\lambda$=850 nm are superimposed at the position of the Bose–Einstein condensate formed in a magnetic trap. This yields a lattice of simple cubic geometry with a lattice constant of $a_{lat}=\lambda/2$=425 nm. After a slow ramp-up of the lattice in 160 ms, the atoms are strongly confined at the lattice sites (potential depth ~$50 E_{rec}$, with $E_{rec}=h^2/2m\lambda^2$) until they are released to free ballistic expansion by switching off all potentials.

Following a time of flight period, the two-dimensional density profile of the cloud is obtained by illuminating it with a resonant laser pulse and projecting the profile of the resulting beam onto a CCD camera. A second image is taken without the atoms in the beam, and the two resulting images are divided to determine the optical density distribution of the cloud. The number of atoms in a column corresponding to a region of the imaging plane can then be deduced from the integrated optical density in that region[24].

**Analysis of images**

In addition to the finite resolution, the camera system adds artefacts to the images owing to optical interference effects of the coherent illumination light and electronic crosstalk during readout of the CCD chip. In our case, the former results in a pattern of vertical stripes and the latter mainly creates a periodic noise with a wavelength of two pixels. As the phase and amplitude of both periodic distortions are not constant, they can not be cancelled by the normalization procedures and appear as periodic fluctuation in the noise correlation plot. Images with high amplitude of such fluctuations (visible outside the atom cloud) are removed from further analysis. The electronic noise is addressed after the determination of the correlation function, by convolving it with a horizontal three-pixel-wide gaussian mask for smoothing.

The correlation function as defined in equation (1) is obtained from a set of images as follows: from each image the autocorrelation function (ACF) is calculated by Fourier-transforming it, taking the absolute square to obtain the power spectral density and Fourier-transforming it back. Averaging the ACF of all images yields the numerator of equation (1), whereas the denominator is obtained by calculating the ACF of the average of all images.

**Theoretical model**

The origin of the correlation peaks can be understood as follows. Calculating the ACF determines the expectation value of the operator $\langle \hat{n}(\mathbf{x}_1,t)\hat{n}(\mathbf{x}_2,t)\rangle = \langle \hat{a}^+(\mathbf{x}_1,t)\hat{a}(\mathbf{x}_1,t)\hat{a}^+(\mathbf{x}_2,t)\hat{a}(\mathbf{x}_2,t)\rangle$ at time $t$, with $\mathbf{x}_1 = \mathbf{x} - \tfrac{1}{2}\mathbf{d}$, $\mathbf{x}_2 = \mathbf{x} + \tfrac{1}{2}\mathbf{d}$. The operators $\hat{a}(\mathbf{x},t)$ at time $t$ after release relate to the on-site operators $\hat{a}(\mathbf{r}_j)$ for the lattice sites $j$ at positions $\mathbf{r}_j$ as

$$\hat{a}(\mathbf{x},t) = \sum_j w(\mathbf{x}-\mathbf{r}_j,t) e^{i\frac{m}{2\hbar t}(\mathbf{x}-\mathbf{r}_j)^2} \hat{a}(\mathbf{r}_j)$$

where $w$ is the expanding wavefunction originally localized to the Wannier function at the site. For the product of Fock states representing the Mott insulator with site occupation $n_i$ at site $i$, one finds

$$\langle \hat{a}^+(\mathbf{r}_k)\hat{a}^+(\mathbf{r}_m)\hat{a}(\mathbf{r}_l)\hat{a}(\mathbf{r}_n)\rangle = n_k n_m \delta_{kl}\delta_{mn} + n_k n_m \delta_{kn}\delta_{lm} \tag{4}$$

where the delta-term introduced through the normal ordering of the operators has been omitted. In the correlation function $C$, the first term in equation (4) will create a constant offset of 1 for large atom number $N$, whereas the second term introduces a spatial dependence in the correlations, leading to:

$$C_{3D}(\mathbf{d}) = C(\mathbf{x}_1 - \mathbf{x}_2) = 1 + \frac{1}{N^2}\sum_{k,l} e^{i\frac{m}{\hbar t}(\mathbf{x}_1-\mathbf{x}_2)\cdot(\mathbf{r}_k-\mathbf{r}_l)} n_k n_l \tag{5}$$

Throughout the discussion, constant offsets of order $1/N$ are neglected compared to 1. For a regular one-dimensional lattice with unity filling and spacing $a_{\text{lat}}$, the sum can then be simplified to $1+\{[\sin^2(\pi N d/l)]/[N^2\sin^2(\pi d/l)]\}$, with $d=x_2-x_1$ and $l=ht/(ma_{\text{lat}})$ analogous to the optical interference created by a regular grating. In the limit of large $N$, this term corresponds to a series of peaks of height 1 and width $l/N$ and converges to:

$$1 + \frac{1}{N}\sum_{j=-\infty}^{\infty} \delta(d/l - j)$$

For a regular three-dimensional system the structure term converges to:

$$C_{3D}(\mathbf{d}) = 1 + \frac{1}{N}\sum_j \delta((\mathbf{d}-\mathbf{p}_j \tfrac{t}{m})/l)$$

where $\mathbf{p}_j$ are the reciprocal three-dimensional lattice momenta. Because the imaging system registers only column densities and has a finite resolution, the operators $\hat{n}(\mathbf{x}_{1,2})$ both have to be convolved with the inverse point spread function (approximated as a gaussian of r.m.s. width $\sigma$) and integrated along the imaging axis before being evaluated. For unity filling this yields a smoothed two-dimensional correlation function:

$$C(\mathbf{d}) = 1 + \frac{1}{4\pi N}\left(\frac{l}{\sigma}\right)^2 \sum_j e^{-\frac{(\mathbf{d}-\mathbf{p}_j\tfrac{t}{m})^2}{4\sigma^2}}$$

The heights of the peaks at the reciprocal lattice momenta therefore scale as $N^{-1}t^2$ for this simple homogeneous case. As indicated in the text, the $N^{-1}$ scaling is modified to $N^{-0.64}$ for our harmonically trapped system by the appearance of Mott domains with filling factor larger than one for higher atom numbers. The prediction for the exponent has been obtained by numerically evaluating the sum in equation (5) using a model distribution of atoms in the lattice sites confined by a global parabolic potential. This distribution is predicted assuming the system can be described in the strongly interacting limit[30] with a local density approximation.

**Acknowledgements** We acknowledge discussions with E. Altman and M. Greiner, as well as financial support by the DFG, AFOSR and the EU under a Marie-Curie Fellowship (F.G.) and a Marie-Curie Excellence grant.

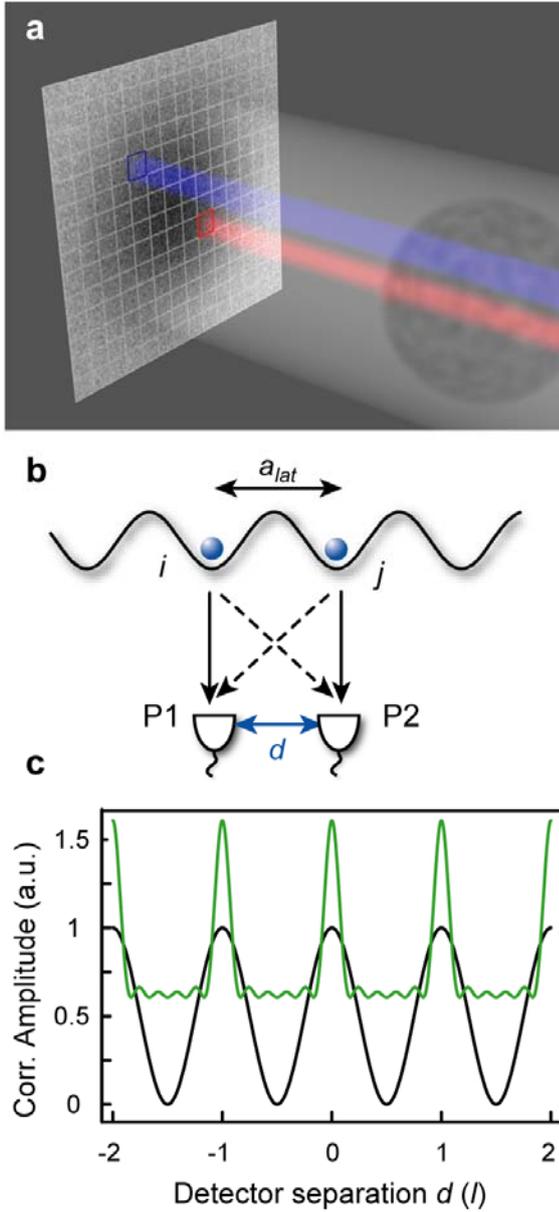

**Figure 1** Illustration of the atom detection scheme and the origin of quantum correlations. **a**, The cloud of atoms is imaged to a detector plane and sampled by the pixels of a CCD camera. Two pixels P1 and P2 are highlighted, each of which registers the atoms in a column

along its line of sight. Depending on their spatial separation *d*, their signals show correlated quantum fluctuations, as illustrated in **b**. **b**, When two atoms initially trapped at lattice sites i and j (separated by the lattice spacing $a_{lat}$) are released and detected independently at P1 and P2, the two indistinguishable quantum mechanical paths, illustrated as solid and dashed lines, interfere constructively for bosons (or destructively for fermions). **c**, The resulting joint detection probability (correlation amplitude) of simultaneously finding an atom at each detector is modulated sinusoidally as a function of *d* (black curve). The multiple wave generalization to a regular array of six sources with the same spacing is shown in green. a.u., arbitrary units.

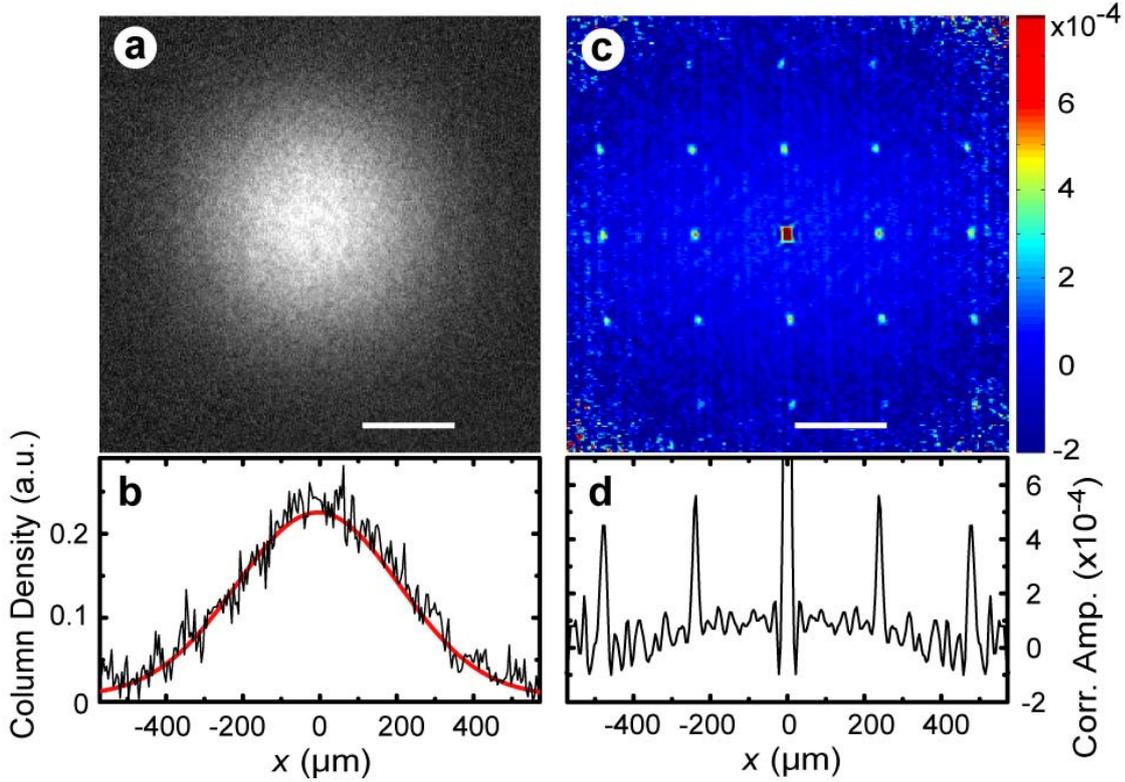

**Figure 2** Single shot absorption image including quantum fluctuations and the associated spatial correlation function. **a**, Two-dimensional column density distribution of a Mott insulating atomic cloud containing $6\times10^5$ atoms, released from a three-dimensional optical lattice potential with a lattice depth of $50E_r$. The white bars indicate the reciprocal lattice scale *l* defined in equation (2). **b**, Horizontal section (black line) through the centre of the image in **a**, and gaussian fit (red line) to the average over 43 independent images, each one similar to **a**. **c**, Spatial noise correlation function obtained by analysing the same set of images, which shows a regular pattern revealing the lattice order of the particles in the trap. **d**, Horizontal profile through the centre of the pattern, containing the peaks separated by integer multiples of *l*. The width of the individual peaks is determined by the optical resolution of our imaging system.

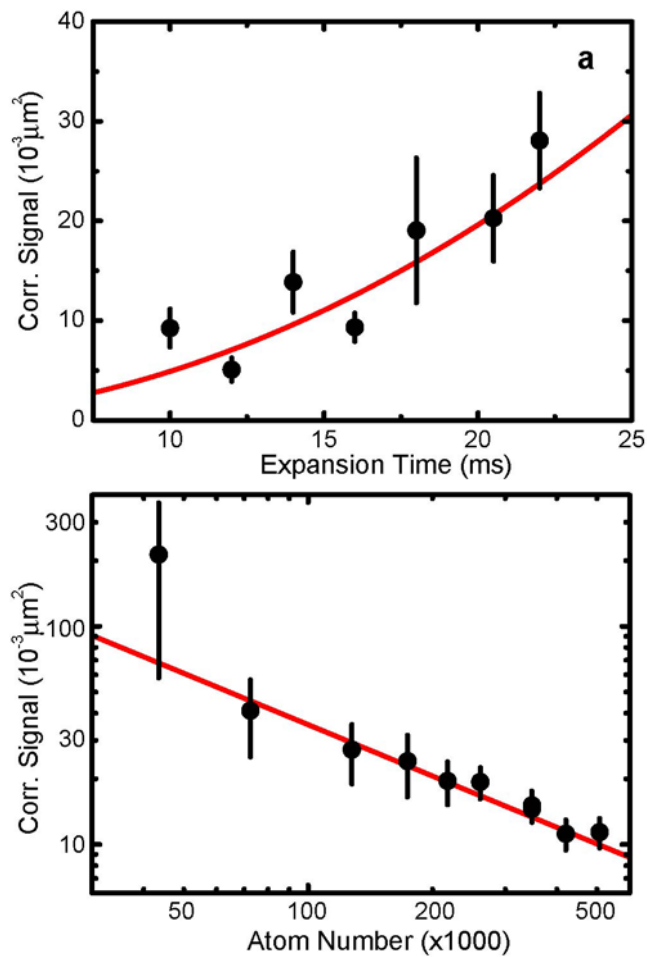

**Figure 3** Correlation signal versus expansion time and atom number. **a**, **b**, Average correlation signal as a function of the time of ballistic expansion (**a**) and the number of atoms *N* loaded in the lattice (**b**). The solid lines denote the result of a simultaneous fit to both data sets to determine the amplitude of the signal and the power law of the decay in **b** (error bars denote root mean square of deviations).